\documentclass{PoS}
\usepackage{graphicx,amsmath,amssymb,dcolumn,bm}

\title{ Probing quark transversity GPDs in diffractive photo- and electroproduction on the deuteron}

\ShortTitle{ Probing quark transversity GPDs in diffractive photo- and electroproduction on the deuteron}

\author{\speaker{W. Cosyn}\\
        Department of Physics and Astronomy, Ghent University, B9000 Ghent, Belgium\\
        E-mail: \email{wim.cosyn@ugent.be}}

\author{B. Pire\\
        CPHT, CNRS, \'{E}cole Polytechnique, I. P. Paris, F-91128 Palaiseau, France\\
        E-mail: \email{bernard.pire@polytechnique.edu}}
        
\author{L. Szymanowski\\
National Centre for Nuclear Research (NCBJ), 02-093 Warsaw, Poland\\
E-mail: \email{Lech.Szymanowski@ncbj.gov.pl}}

\abstract{Transversity generalized parton distributions (GPDs) can be probed in diffractive electro- and photoproduction of two vector mesons on a hadron in kinematics where the two vector mesons are separated by a large rapidity gap.  We report on calculations for this process in the case of coherent $\rho^0-\omega$ meson production on a deuteron target.  Our cross section results show that an electron-ion collider with deuteron beams and forward detectors could probe deuteron transversity GPDs. }

\FullConference{XXVII International Workshop on Deep-Inelastic Scattering and Related Subjects - DIS2019\\
		8-12 April, 2019\\
		Torino, Italy}

\begin{document}

%\section{...}

Generalized parton distributions (GPDs) encode non-perturbative aspects of QCD in hadrons and appear in the decomposition of off-forward quark and gluon correlators of these hadrons.  Recently, the transversity GPDs for spin-1 hadrons were introduced~\cite{Cosyn:2018rdm} and their polynomiality properties were studied~\cite{Cosyn:2018thq}.  In this proceedings, we report on progress in the phenomenology of these transversity GPDs in the case of the deuteron ($D$).

Due to QCD factorization theorems, GPDs appear in the amplitudes of hard exclusive processes.  In the case of quark transversity GPDs, however, the case is complicated due to their chiral-odd nature.  The most natural process -- hard exclusive meson production (HEMP) with transverse meson polarisation -- does not permit the appearance of the transversity GPDs at leading twist~\cite{Diehl:1998pd,Collins:1999un}.  Hence, to probe these GPDs one needs to include higher twist contributions~(for instance in $\pi^0$ production\cite{Ahmad:2008hp,Goloskokov:2011rd}) or consider processes with more particles in the final state~\cite{Boussarie:2016qop}.

Here, we consider the diffractive electro- or photoproduction of two vector mesons on a hadron, where the two mesons are separated by a large rapidity gap:
\begin{equation}\label{eq:reaction}
\gamma^{(*)} h(p) \rightarrow V_{1}^0 V_2 h'(p')\,.
\end{equation}

  This process was first considered for double $\rho$-meson production on the proton in Refs.~\cite{Ivanov:2002jj,Enberg:2006he}.  For detailed expressions and derivation of the amplitude and cross section, we refer to these references. Here, we summarize the main properties and conclusions.

\begin{figure}[ht]
\begin{center}
\includegraphics[width=.45\textwidth]{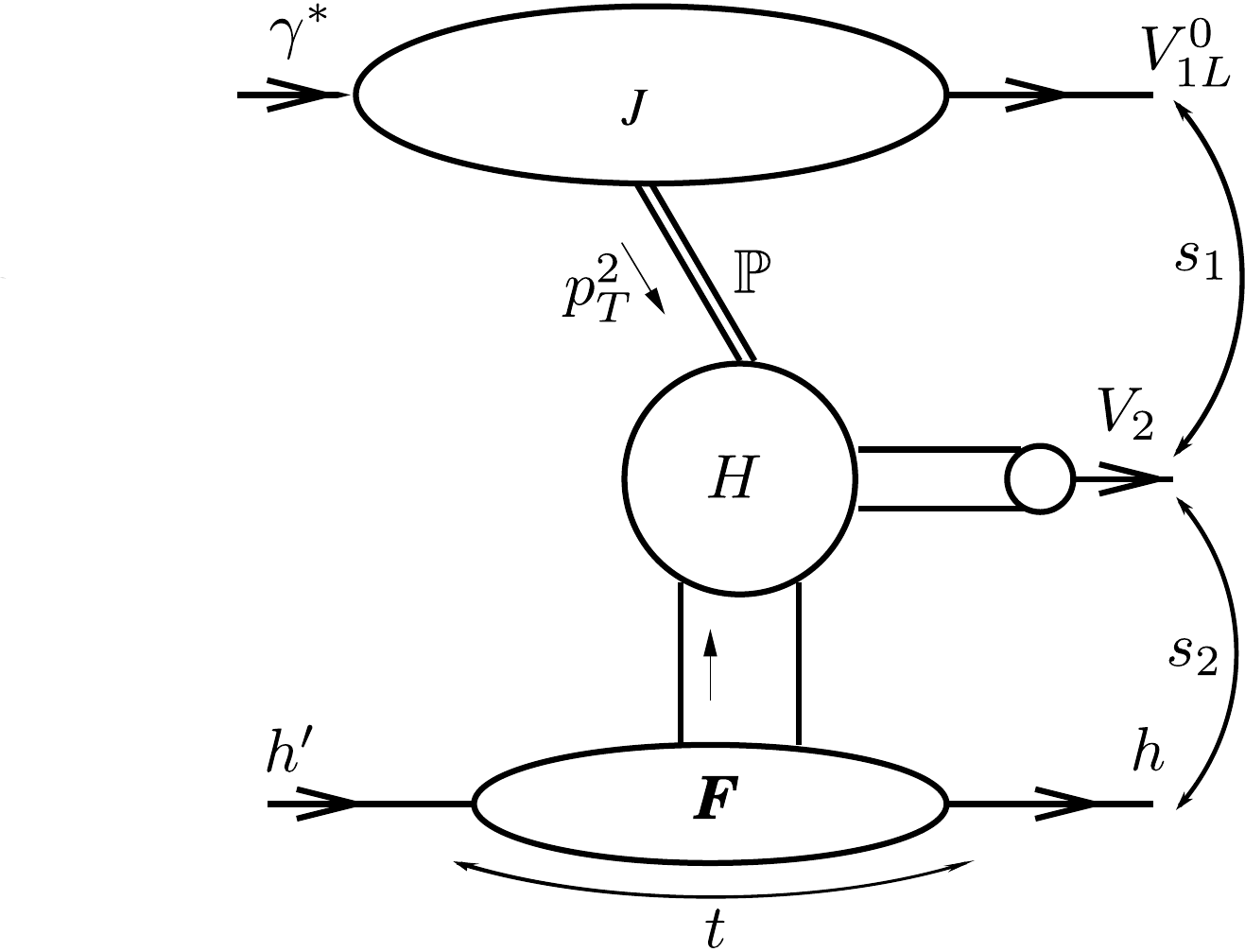}
\caption{Diagrammatic representation of QCD factorization for the diffractive two vector meson production amplitude.  See text for kinematic variables.}
\label{fig:diagram}
\end{center}
\end{figure}

The kinematics considered are: i) invariant mass $s_1$ of the two vector mesons $(V_{1L}^0-V_2)$  is large and of the order of the total invariant mass of the reaction $s$; ii) $(V_2-h')$ invariant mass $s_2$ is smaller than $s_1$ but of the order of $p_T^2$ (see below), i.e. of a hard scale.  In these kinematics and at leading twist (LT), the amplitude of the process (\ref{eq:reaction}) can be written as the convolution of an impact factor $J^{\gamma^{(*)}_{L/T} V_{1L}^0}$ (which describes the photon - vector meson $V_1$ transition by exchange of two gluons, seen as the Born approximation of the hard pomeron), and the amplitude of the pomeron - hadron interaction $\mathbb{P} h \rightarrow V_2 h'$.  The virtuality of the pomeron corresponds to the hard scale $p_T^2$, which means the second subprocess can be described using a GPD correlator for the hadron and a distribution amplitude for $V_2$, see Fig.~\ref{fig:diagram}.  The polarization of meson $V_2$ then determines whether the vector (longitudinal polarization) or transversity GPDs (transverse) of the hadron enter. Let us stress that the target does not need to be polarized to access transversity GPDs.

The hard scattering amplitude of the process has been computed at leading order (LO) in the collinear approximation in Ref.~\cite{Ivanov:2002jj}.  Because of the charge conjugation properties of the Pomeron and vector mesons, gluon GPDs do not contribute to the process. Six diagrams contribute (corresponding to the two exchanged gluons coupling to different quark lines in the upper and lower part of the diagram). It was shown that the total amplitude has no pinch singularities, and end-point singularities are regularized by the distribution amplitudes of the mesons and the impact factor.  Hence, QCD factorization holds at leading order and the GPD framework can be applied.  It was also shown that the GPDs are probed in the ERBL region ($|x|<|\xi|$), where $x$ is the average momentum fraction of the two quarks involved in the hadron GPD correlator and $\xi$ is the standard skewness variable.  This means the process probes a $q\bar{q}$ pair in the hadron, which turns into the $V_2$ meson.  The differential cross section in these LT and LO approximations $\frac{\text{d}\sigma}{\text{d}t \text{d}p^2_T \text{d}\xi}$ (where $t=(p-p')^2$ is the momentum transfer squared to the hadron) turns out to be independent of total invariant mass $s$~\cite{Enberg:2006he}.

The extension to the case of coherent production on the deuteron is straightforward: due to the isoscalar nature of the deuteron we have to consider $\omega$ or $\phi$-meson production in the second subamplitude $\mathbb{P}D \rightarrow (\omega / \phi)  D'$ and consequently the isoscalar deuteron GPDs enter in the amplitude.  Details will be presented elsewhere~\cite{CPS}.

\begin{figure}[h]
\begin{center}
\includegraphics[width=.48\textwidth]{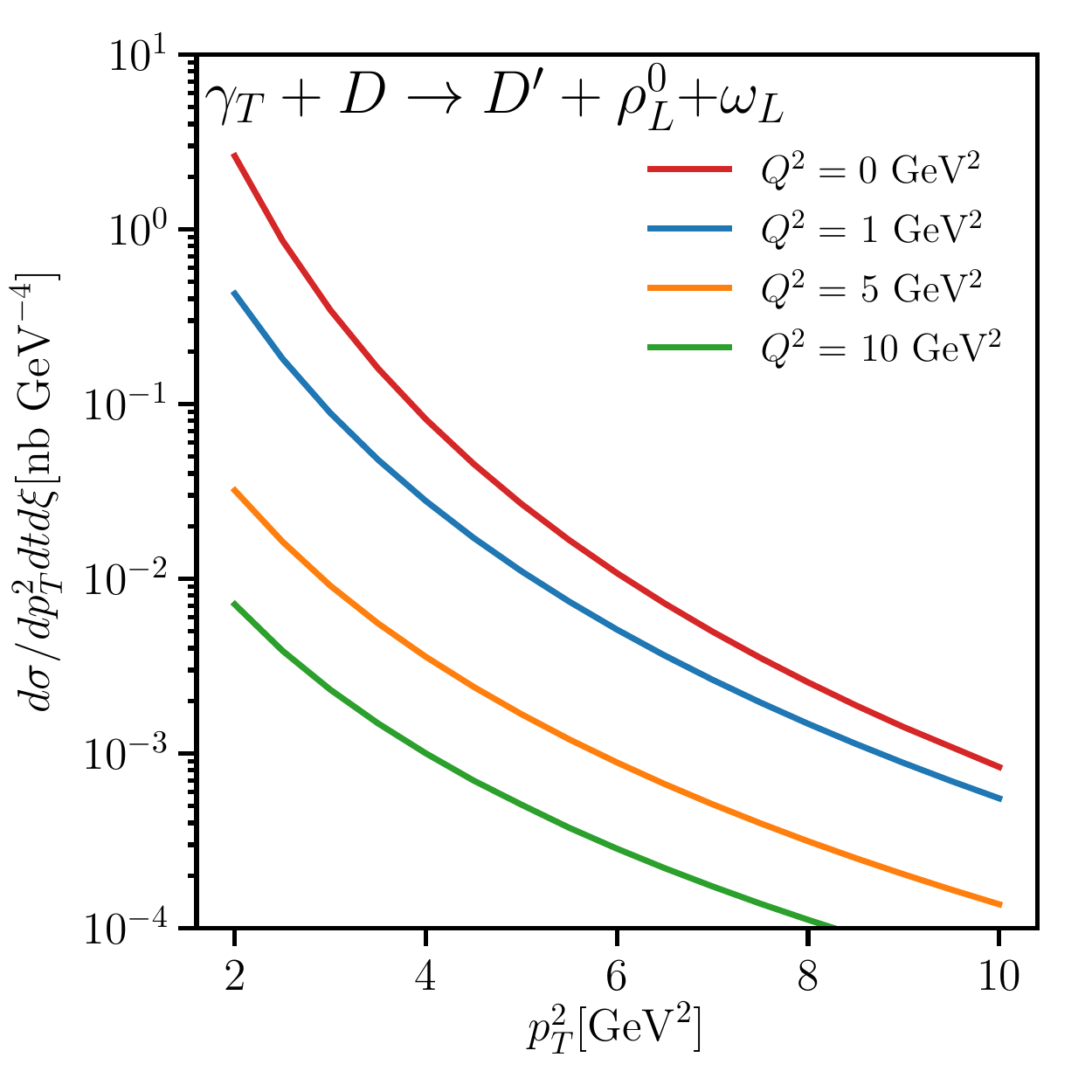}
\includegraphics[width=.48\textwidth]{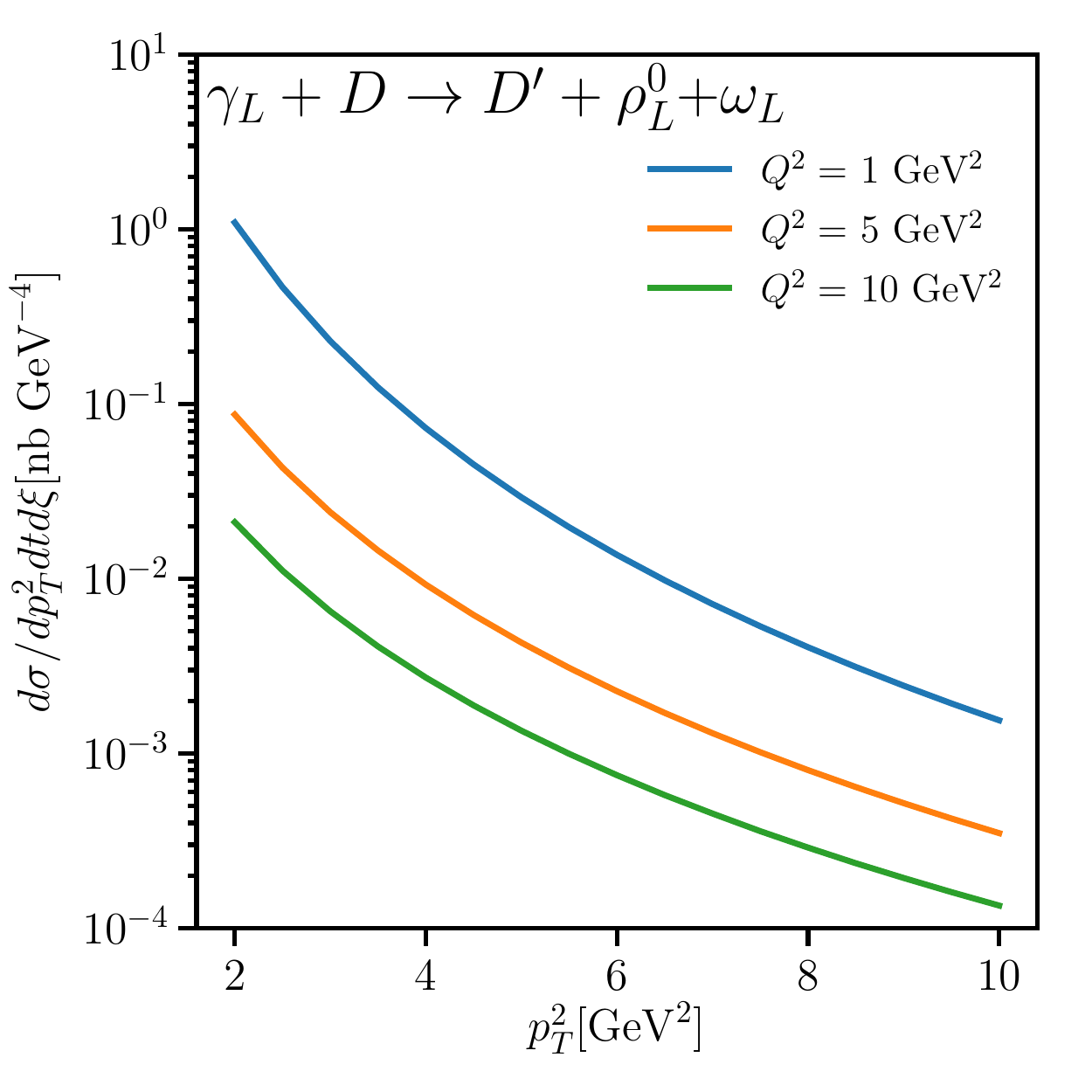}

\caption{Differential cross section of coherent diffractive two vector meson  ($\rho^0_L$ and $\omega_L$) production on the deuteron as a function of pomeron virtuality for different values of the photon virtuality; $\xi=0.15$ and $t=t_\text{min}=-0.33~\text{GeV}^2$.  Left panel shows transverse photon polarization (and includes photoproduction), right panel shows  longitudinal photon polarization.}
\label{fig:omegaL}
\end{center}
\end{figure}

\begin{figure}[h]
\begin{center}
\includegraphics[width=.48\textwidth]{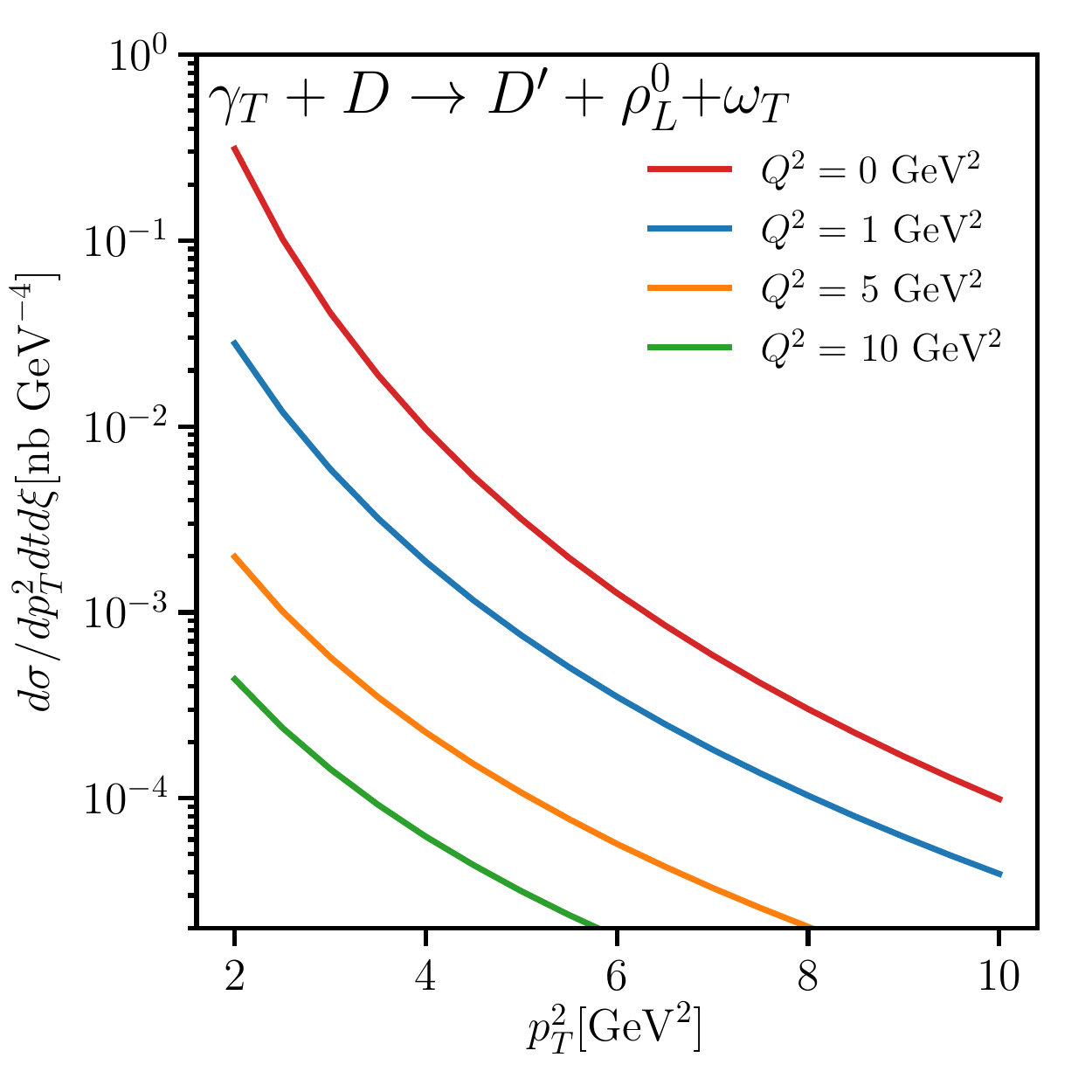}
\includegraphics[width=.48\textwidth]{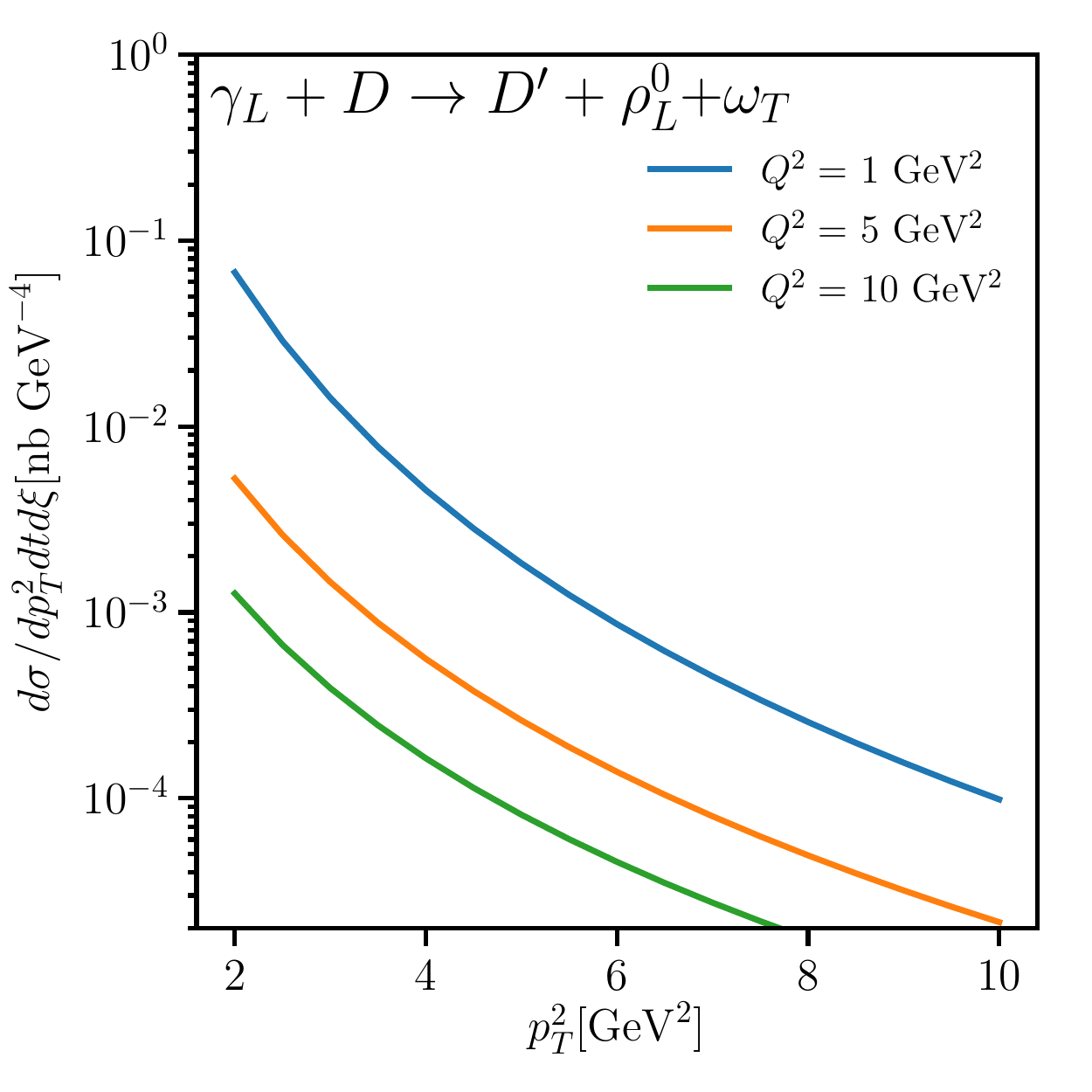}

\caption{As Fig.~\ref{fig:omegaL} but for transverse omega polarization in the final state.}
\label{fig:omegaT}
\end{center}
\end{figure}

To compute the cross section for the reaction on the deuteron, we need a framework to compute the required deuteron GPDs.  We use a convolution model, described in more detail in Refs.~\cite{Cano:2003ju,Cosyn:2018rdm}.  In this convolution model, we only consider the dominant nucleon-nucleon Fock state of the deuteron and limit ourselves to the leading order impulse approximation where one of the nucleons acts as a spectator, and the bilocal parton correlator operator acts on the other nucleon.  Using methods of light-front quantization, nucleon and nuclear structure are separated and the GPD correlator of the deuteron is written as the convolution of the $np$ component of two light-front deuteron wave functions and a nucleon GPD correlator.  The latter is parameterized using isoscalar proton GPDs. One drawback of this basic convolution model is that it violates the polynomiality properties of the deuteron GPDs as the truncation of the Fock space in the convolution violates Lorentz invariance.  Extensions that remedy this will be a topic of future study.

In Figs.~\ref{fig:omegaL} and \ref{fig:omegaT}, we show differential cross sections for $\rho^0_L$ and $\omega$ production on the deuteron as a function of the pomeron virtuality $p_T^2$.  The following ingredients enter in the calculations.  The AV18 deuteron radial wave functions~\cite{AV18} enter in the deuteron light-front wave function used in the convolution.  The PARTONS framework~\cite{Berthou:2015oaw} is used for the vector nucleon GPDs, and more specifically the MMS13 model~\cite{Mezrag:2013mya} implemented therein.  For the nucleon transversity GPDs, we use the model of Ref.~\cite{Goloskokov:2011rd}.  We have evaluated the cross section at the minimum value of $-t_\text{min}$, with  $t_\text{min}=-\frac{4M_h^2\xi^2}{1-\xi^2}$ (where $M_h$ is the hadron mass).  On the one hand, this yields the largest cross sections, and on the other hand it limits the number of GPDs contributing to the amplitude.

First, Fig.~\ref{fig:omegaL} shows cross sections for a longitudinally polarized $\omega$ meson, resulting in the deuteron vector GPDs~\cite{Berger:2001zb} contributing to the process.  Fig.~\ref{fig:omegaT} shows results with a transversely polarized $\omega$ meson, meaning deuteron transversity GPDs contribute.  Both calculations result in a larger photoproduction cross section than for electroproduction (where the additional reduction through the virtual photon flux is not included in the plots).  The results with a longitudinal polarized photon yield larger cross sections than the transverse photon ones at equal $Q^2$ value.  Comparing the different polarizations for the $\omega$ meson, we observe that the process probing deuteron transversity GPDs results in smaller cross sections.

To conclude, we show that diffractive double vector meson production on the deuteron results in sizeable cross sections, which means deuteron transversity GPDs could be probed at an electron-ion collider~\cite{Boer:2011fh,Accardi:2012qut} with forward detectors and sufficient angular coverage to determine the polarization of the produced vector mesons.  More detailed studies will be covered in an upcoming article~\cite{CPS}.

\section*{Acknowledgments}
This project has received funding from the European Union's Horizon 2020 research and innovation programme under grant agreement No 824093. L.S. is supported by the grant 2017/26/M/ST2/01074 of the National Science Center in Poland. L.S. also thanks the French LABEX P2IO, the French GDR QCD and the LPT for support.

\bibliographystyle{JHEP}
\bibliography{../bibtexall.bib}

\end{document}